\pdfoutput=1
\documentclass{article}
\usepackage[utf8]{inputenc} 
\usepackage[T1]{fontenc}
\usepackage{hyperref}
\usepackage{url}      
\usepackage{setspace}
\usepackage[pdftex]{graphicx}
\usepackage{dcolumn}
\usepackage[margin=1in]{geometry}
\usepackage{amsmath}
\usepackage{graphicx}
\usepackage{float}
\usepackage{tabularx}
\usepackage[sorting=none]{biblatex}
\bibliography{reference.bib}
\usepackage[mathscr]{eucal}
\usepackage{multirow}
\usepackage{booktabs}

\begin{document}

\title{Inflation in f(R,T) gravity with Double-Well potential}

\author{ Biswajit Deb \footnote{Electronic Address:biswajitdeb55@gmail.com} , Atri Deshamukhya \footnote{Electronic Address: atri.deshamukhya@gmail.com}\\
Department of Physics, Assam University, Silchar, India }

\date{}
\maketitle

\begin{abstract}
In this piece of work, we studied the inflation in the context of ${f(R,T)}$ theory of gravity. We assumed the functional form of ${f(R,T)}$ to be $R+ 16 \pi G \lambda T$, where R is the Ricci scalar, T is the trace of the Energy-Momentum tensor and $\lambda$ is the model parameter. The cosmological observable parameters like scalar spectral index $n_s$ and tensor-to-scalar ratio $r$ are estimated for Double-Well potential. We found that for $\lambda = 150$, $n_s$ and $r$ are in good agreement with Planck 2018 data. Further, considering the vacuum expectation value in Double-Well potential to be Planckian, we observed the admissible range of model parameter to be $145 < \lambda < 222$ for which this model remains consistent with Planck 2018 data.

\end{abstract}

\section{Introduction}	

The current observational data from COBE, WMAP, Planck confirms that the universe must had gone through a rapid exponential phase of expansion called inflation in its primordial stage \cite{r1,r2,r3}. This inflationary scenery is necessary to have a perfect solution for horizon and flatness problem if we consider general relativity is the correct description of the universe. Inflation also provides the seeds for density perturbation and large scale structure of the universe. The concept of inflation was originally proposed by Alan Guth \cite{r4} and subsequently developed by other scientists like Starobinsky \cite{r5}, Liddle \cite{r6}, Riotto \cite{r7} etc.\\ \\
In this scenario it is assumed that a scalar field called inflaton rolls down its potential to give rise to an inflationary era \cite{r4}. Several types of potential have been used to describe inflation, however most of them have been constrained and excluded using scale invariant power spectrum of scalar perturbation data from Planck as well as WMAP \cite{r8, r9}. \\ \\
Apart from scalar field theory, inflation can also be realised considering large scale modification of general relativity where the Einstein-Hilbert action is modified and higher order terms of Ricci scalar are added to gravitational Lagrangian. The modified theories of gravity gained interest because Einstein's General Relativity can't explain all the evolutionary phases simultaneously. Besides this, GR fails to explain the dark sector of the universe. As a result, cosmologist started studying the different corrections to GR and their implication on realising the observed universe. f(R) theory or the $R^2$ inflation is the first of its kind proposed by Starobinsky in this genre \cite{r5}. Besides f(R) theory, inflation has also been studied in  f($\mathscr{T}$) gravity \cite{r10}, $f(G)$ gravity \cite{r11} theories where $\mathscr{T}$ , G denotes the Torsion and Gauss-Bonnet scalars respectively. \\ \\
The ${f(R,T)}$  gravity was proposed primarily by Harko et al. where R is the Ricci scalar and T is the trace of the energy-momentum tensor \cite{r12}. ${f(R,T)}$ theory has been studied in different topics of cosmology such as inflation \cite{r13, r14, r15, r16}, dark energy \cite{r17, r18, r19, r20, r21}, dark matter \cite{r22}, wormhole \cite{r23, r24, r25, r26, r27, r28, r28a, r28b}, gravitational waves \cite{r29, r30, r31}, baryogenesis \cite{r32}, density perturbation \cite{r33}, scalar field models \cite{r34, r35}, anisotropic model \cite{r36, r37}, bouncing cosmology \cite{r38, r39}, big-bang neucleosysnthesis \cite{r40},  brane world \cite{r41, r42}, pulsars \cite{r43, r44}, white dwarfs \cite{r45}. Further, the energy conditions \cite{r46, r47, r48, r49} and junction conditions \cite{r50} in ${f(R,T)}$ gravity have also been studied. The first study of inflationary dynamics in the framework of ${f(R,T)}$ gravity was carried out by Snehasish et al. where they studied inflation from perfect fluid and quadratic potential \cite{r13}. Subsequently, Mauricio showed that in case of monomial power law potentials, no corrections due to ${f(R,T)}$ have been induced whereas in Natural and Hilltop potential, one of the parameter is trivially modified \cite{r14}. Further, they showed that Starobinsky model in ${f(R,T)}$ predicts consistent result with the observational data \cite{r14}. Chen et al. studied the non-minimal coupling of R and T in ${f(R,T)}$ gravity and showed that the presence of mixing term RT in the theory can rescue Chaotic and Natural inflation from rejection. \cite{r16}  \\ \\
With the recent development in CMB anisotropy measurements, the range of the observable parameters like spectral indices and tensor-to-scalar ratio have been constrained \cite{r9}. As a result, most of the inflationary potentials have been rejected \cite{r8, r9}. In ${f(R,T)}$ models, the presence of  trace term T in the theory may show some exciting result and leave scope for the rejected potentials to be consistent with observational data. Further, there are potentials whose vacuum expectation value ({\it{vev}}) shows Super Planckian scales in order to match the current CMB data \cite{r51}. These Super Planckian scales are problematic from the Particle Physics aspects since it doesn't permit Inflation scale to be Super Planckian \cite{r52}. However, modified gravity theories might rescue these potentials from rejection and set their {\it{vev}} at Sub-Planckian ranges. In this piece of work we have taken one such potential viz. Double-Well potential and will study the impact of trace term correction on the model results. \\ \\
The paper has been organised as follows: In section 2, we review ${f(R,T)}$ gravity theory in brief and in section 3 the slow-roll dynamics in {\it{f(R,T)}} gravity has been studied. In section 4, we study inflationary scenario with double-well potential in the background of ${f(R,T)}$ gravity. In section 5, we present our conclusion. Here, we will use natural system of unit with $c= \hbar  =1$ and the (-,+,+,+) sign convention for the metric tensor.


\section{{\it{f(R,T)}} gravity in nutshell}
The action in ${f(R,T)}$ gravity theory as proposed by Harko \cite{r12} is given by,

\begin{equation}
    S = \int \left[\frac{f(R,T)}{16 \pi G} + L_m  \right] \sqrt{-g} d^4x 
\end{equation}
 where ${f(R,T)}$ is an arbitrary function of Ricci scalar R and the trace of the energy-momentum tensor $T_{\alpha\beta}$, $L_m$ is the matter Lagrangian density, g is the metric determinant and G is the Newtonian gravitational constant. On variation of the action with respect to the metric, we obtain the ${f(R,T)}$ gravity field equation as,
 \begin{equation}
    f_R (R,T) R_{\alpha\beta} - \frac{1}{2} g_{\alpha\beta}f(R,T) + [g_{\alpha\beta} \nabla_\sigma \nabla^\sigma - \nabla_\alpha \nabla_\beta] f_R (R,T) =  8\pi G T_{\alpha\beta} - f_T (R,T) (T_{\alpha\beta} + \Theta_{\alpha\beta})
\end{equation}
where we have denoted $f_R (R,T)= \frac{\partial f(R,T)}{\partial R}$ , $f_T (R,T)= \frac{\partial f(R,T)}{\partial T}$ and defined $T_{\alpha\beta}$ and $\Theta_{\alpha\beta}$ as,
\begin{equation}
     T_{\alpha\beta} = g_{\alpha\beta}L_m - 2 \frac{\delta L_m}{\delta g^{\alpha\beta}}
\end{equation}
\begin{equation}
    \Theta_{\alpha\beta}= g^{\mu\nu} \frac{\delta T_{\mu\nu}}{\delta g^{\alpha\beta}} = -2 T_{\alpha\beta} + g_{\alpha\beta}L_m - 2 \frac{\delta^2 L_m}{\delta g^{\alpha\beta} \delta g^{\mu\nu}}
\end{equation}
This term $\Theta_{\alpha\beta}$ plays a crucial role in ${f(R,T)}$ gravity. Since it contains matter Lagrangian $L_m$, depending on the nature of the matter field, the field equation for ${f(R,T)}$ gravity will be different. Besides this the functional form of ${f(R,T)}$ will also change the field equation. Thus, the field equation in ${f(R,T)}$ gravity depends both on the nature of matter field and choice of the function $f$. \\ \\
Harko proposed four different forms of the function ${f(R,T)}$ in his paper\cite{r12}, viz.:
\begin{itemize}
    \item $f(R,T) = R + 2 f(T)$
    \item $f(R,T) = f_1(R) + f_2(T)$
    \item $f(R,T) = f_1(R) + f_2(R)f_3(T)$
    \item $f(R,T^{\phi}) = R + f(T^{\phi})$, where $\phi$ is a self interacting scalar filed.
\end{itemize}
Apart from the above forms, Mauricio\cite{r14} proposed two other forms of ${f(R,T)}$ which are as follows:
\begin{itemize}
    \item $f(R,T) = R + \alpha RT$
    \item $f(R,T) = R + \alpha e^{RT}$, $\alpha$ is a constant.
\end{itemize}
In this work, we have considered the simplest form of ${f(R,T)}$ which is $f(R,T) = R + 16 \pi G \lambda T$, where $\lambda$ is the model parameter. In this particular case the action becomes,
\begin{equation}
    S = \int \left[\frac{R}{16 \pi G} +  \lambda T + L_m \right] \sqrt{-g} d^4x 
\end{equation}
and the field equation will take the following form,
\begin{equation}
    R_{\alpha\beta} - \frac{1}{2}g_{\alpha\beta}R = 8 \pi G T_{\alpha\beta}^{(eff)}
\label{5}
\end{equation}
where $T_{\alpha\beta}^{(eff)}$ is the effective stress-energy tensor given by,
\begin{equation}
    T_{\alpha\beta}^{(eff)}= T_{\alpha\beta} - 2 \lambda( T_{\alpha\beta} - \frac{1}{2}T g _{\alpha\beta} + \Theta_{\alpha\beta})
\label{2}
\end{equation}
It is clear that when $\lambda=0$, above field equation immediately reduces to Einstein's general field equation.


\section{Inflation in {\it{f(R,T)}} gravity}
To understand the cosmological implications of ${f(R,T)}$ gravity, let us assume the Friedmann-Lemaitre-Robertson-Walkar (FLRW) metric in spherical coordinate,
\begin{equation}
    ds^2= - dt^2 + a(t)^2 \left[\frac{dr^2}{1-kr^2}+r^2 (d\theta^2 + \sin^2\theta d\phi^2)\right]
\end{equation}
where a(t) is the scale factor with $t$ being the cosmic time. The constant k is the spatial curvature with value, +1, -1, and 0 which corresponds to close, open and flat universe respectively. The current observation shows that the present universe is spatially flat and hence we will assume k=0. \\ \\
Considering the above metric, various components of Ricci tensor are as follows \cite{r53}:
\begin{align}
    R_{00} &= - 3 \frac{\dot a}{a}, \nonumber \\
    R_{ij} &= \left[\frac{\ddot a}{a} + 2 \left(\frac{\dot a}{a}\right)^2 \right]g_{ij}, \nonumber \\
    R &= 6 \left[ \frac{\ddot a}{a} + \left(\frac{\dot a}{a}\right)^2 \right]
\end{align}
In order to describe the inflation, we have to introduce a homogeneous scalar field $\phi$ minimally coupled to gravity called inflaton \cite{r54}. Then the matter Lagrangian will take the form,
\begin{align}
    L_m &= -\frac{1}{2}g^{\alpha\beta}\partial_{\alpha}\phi \partial_{\beta}\phi - V(\phi) \nonumber \\
    &= \frac{1}{2}\dot \phi^2 - V(\phi)
\label{1}
\end{align}
where $V(\phi)$ is the potential whose form will be discussed later on. Then the energy-momentum tensor takes the form,
\begin{equation}
    T_{\alpha\beta} = \partial_{\alpha}\phi \partial_{\beta}\phi + g_{\alpha\beta} [\frac{1}{2}\dot \phi^2 - V(\phi)]
\end{equation}
From this we can calculate the energy density $\rho$ and pressure p component of $T_{\alpha\beta}$ as,
\begin{equation}
    T_{00} = \frac{1}{2}\dot\phi^2 + V(\phi) = \rho
\end{equation}
\begin{equation}
    T_{ij} = \left[\frac{1}{2}\dot\phi^2 - V(\phi)\right]g_{ij} = p g_{ij}
\end{equation}
Now we will evaluate the different components of the tensor $\Theta_{\alpha\beta}$. For the matter Lagrangian given in Eq. (\ref{1}), the tensor $\Theta_{\alpha\beta}$ becomes,
\begin{equation}
    \Theta_{\alpha\beta} = -2 \partial_\alpha \phi \partial_\beta\phi - g_{\alpha\beta}\left[\frac{1}{2}\dot\phi^2 - V(\phi)\right]
\end{equation}
and its components are as follow:
\begin{align}
    \Theta_{00} &= - T_{00} - \dot\phi^2; \nonumber \\
    \Theta_{ij} &= - T_{ij}
\end{align}
We will now proceed to compute the components of effective energy-momentum tensor defined in Eq. (\ref{2}),
\begin{align}
    T_{00}^{(eff)} &= \frac{1}{2}\dot\phi^2 (1+ 2\lambda) + V(\phi) (1+ 4\lambda) = \rho^{(eff)}
    \label{3}
\end{align}
\begin{align}
    T_{ij}^{(eff)} &= \frac{1}{2}\dot\phi^2 (1+ 2\lambda) - V(\phi) (1+ 4\lambda) g_{ij} = p^{(eff)}g_{ij}
    \label{4}
\end{align}
From Eq. (\ref{3}) and (\ref{4}) we can derive the effective equation of state as the ratio of effective pressure and energy density,
\begin{equation}
    w^{(eff)} = \frac{p^{(eff)}}{\rho^{(eff)}} = \frac{\dot\phi^2 (1+ 2\lambda) - 2V(\phi) (1+ 4\lambda)}{\dot\phi^2 (1+ 2\lambda) + 2V(\phi) (1+ 4\lambda)}
\end{equation}
Now, the 00 component of field Eq. (\ref{5}) yields,
\begin{equation}
    H^2 = \frac{8\pi G}{3}\left[\frac{\dot\phi^2}{2} (1+ 2\lambda) + V(\phi) (1+ 4\lambda)\right]
\end{equation}
which is known as the modified Friedmann first equation. Here  the Hubble parameter is defined as $H=\frac{\dot a}{a}$. Similarly, computation of the ij component yields,
\begin{equation}
    \frac{\ddot a}{a} = - \frac{8\pi G}{3} \left[\dot\phi^2 (1+ 2\lambda) + V(\phi) (1+ 4\lambda)\right]
    \label{6}
\end{equation}
This is called the modified acceleration equation.

From this, we can obtain the expression for $\dot H$ as,
\begin{equation}
    \dot H = \frac{\ddot a}{a} - H^2 = - \frac{8\pi G}{2} (p^{eff} + \rho^{eff})= - 4\pi G \dot\phi^2(1+ 2\lambda)
    \label{7}
\end{equation}
Both the above Eq. (\ref{6}) and (\ref{7}) are known as modified Friedmann second equation.
Now, using Eq. (\ref{3}) and (\ref{4}) we can derive the continuity equation or the modified Klein-Gordon equation as,
\begin{equation}
    \ddot \phi (1+ 2\lambda) + 3H\dot \phi (1+ 2\lambda) + \frac{dV}{d\phi} (1+ 4\lambda) = 0
\end{equation}
Now,to check whether slow roll inflation is viable in this model, the standard slow-roll parameters need to be computed for this model. One of the standard slow-roll parameter denoted by $\epsilon$  \cite{r54, r55} in terms of parameters of this model are found as: ,
\begin{equation}
    \epsilon= - \frac{\dot H}{H^2} = \frac{3 \dot\phi^2 (1+ 2\lambda)}{\dot\phi^2 (1+ 2\lambda) + 2 V(\phi) (1+ 4\lambda)}
    \label{a}
\end{equation}
Accelerated expansion occurs in slow roll regime if $\epsilon << 1$, that is when the potential energy of the inflaton dominates over the kinetic energy. Under this condition the inflaton rolls down slowly and hence the name, slow-roll condition \cite{r54}. In terms of current model parameters, the first slow roll condition reads as,
\begin{equation}
    \dot\phi^2 (1+ 2\lambda) << V(\phi) (1+ 4\lambda)
    \label{8}
\end{equation}
Again, the accelerated expansion must sustain for a long period of time to achieve the desired amount of e-folds and hence we need another condition to be imposed. In terms of our model parameters this reads as \cite{r54},
\begin{equation}
    \ddot\phi (1+ 2\lambda) << | 3 H \dot \phi (1+ 2 \lambda)|
    \label{9}
\end{equation}
This defines the second slow-roll condition $ |\eta| << 1 $,  \cite{r54, r55} as:
\begin{equation}
    \eta= - \frac{\ddot\phi}{H \dot \phi}
    \label{b}
\end{equation}
The parameters defined in Eq. \ref{a} and \ref{b} are often called Hubble slow-roll parameters \cite{r54}. These parameters can also be defined in terms of the potential and are called potential slow-roll parameters $\epsilon_{V}$ and $\eta_{V}$ \cite{r54, r55}. \\ \\
We  obtain the potential slow-roll parameters in the proposed model  in terms of reduced Planck mass $M_P^2= \frac{1}{8 \pi G}$, where $M_P= 2.4 \times 10^{18}  GeV$ as \cite{r55},
\begin{align}
    \epsilon_V &= \frac{1}{1+ 2 \lambda} \frac{M_P^2}{2} \left(\frac{V'}{V}\right)^2 \nonumber \\
    \eta_V &=  \frac{1}{1+ 2 \lambda} M_P^2 \frac{V''}{V}
    \label{12}
\end{align}

From the above expressions, it is clear that the affect of the correction due to f(R,T) chosen has been induced in the slow-roll parameters. If $\epsilon_V, |\eta_V| << 1$ holds in a region of space, then slow roll inflation can happen \cite{r54}. It may be noted that the slow-roll inflation ends as soon as either of the slow roll parameters becomes of the order of unity.\\ \\
Another important quantity to be studied in any inflationary model is the amount of inflation that has taken place. It is quantified by the no of e-folding  that has occurred before the end of inflation and is defined by the natural logarithm of the scale factor as,
\begin{equation}
    N = \ln a = \int_{\phi_i}^{\phi_{end}} \frac{H}{\dot \phi} \,d\phi\ 
\end{equation}
where $\phi_{end}$ is the value of the inflaton field at the end of the inflation. Under slow-roll approximation, the number of e-folding becomes,
\begin{equation}
    N= \frac{1+2\lambda}{M_P^2} \int_{\phi_{end}}^{\phi_i} \frac{V(\phi)}{V'(\phi)} \,d\phi\ 
    \label{13}
\end{equation}
The spectral indices and tensor-to-scalar ratio in terms of the slow-roll parameters can be written as \cite{r54, r56},
\begin{align}
    n_s - 1 &= \frac{d \ln (\Delta_s^2)}{d \ln (k)} = - 6\epsilon_V + 2 \eta_V  \nonumber \\
    n_T &= \frac{d \ln (\Delta_T^2)}{d \ln (k)} = - 2\epsilon_V \nonumber \\
    r &= 16 \epsilon_V
\end{align}
where $ \Delta_s$ and $\Delta_T$ are the dimensionless scalar and tensor power spectrum respectively.


\section{Inflation from Double-Well potential}
Double-Well potential popularly known as "Maxican hat" potential was first introduced by Goldstone \cite{r57} and is given by,
\begin{equation}
   V(\phi) = M^4 \left[\left(\frac{\phi}{\phi_0} \right)^2 - 1 \right]^2
\end{equation}
where M is the mass scale, $\phi_0$ is the vacuum expectation value.  In case of inflation this potential was used to construct the scenario of topological inflation \cite{r58, r59}. The parameter $\phi_0$  sets the value at which fluctuation proceeds and it can vary over a wide range from GUT symmetry breaking scale ($10^{15} GeV$) to  Super Planckian scale in order to match the predictions of the model with CMB data \cite{r51}. A dimension-less parameter $x=\phi/\phi_0$ has been introduced for convenience of computation. Fig.1 shows the variation of the potential in terms of the dimensionless parameter $x$. The inflation proceeds from left to right in the region $x < 1$.
\vspace*{-2mm}

\begin{figure}[H]
\centerline{\includegraphics[width=0.5 \textwidth] {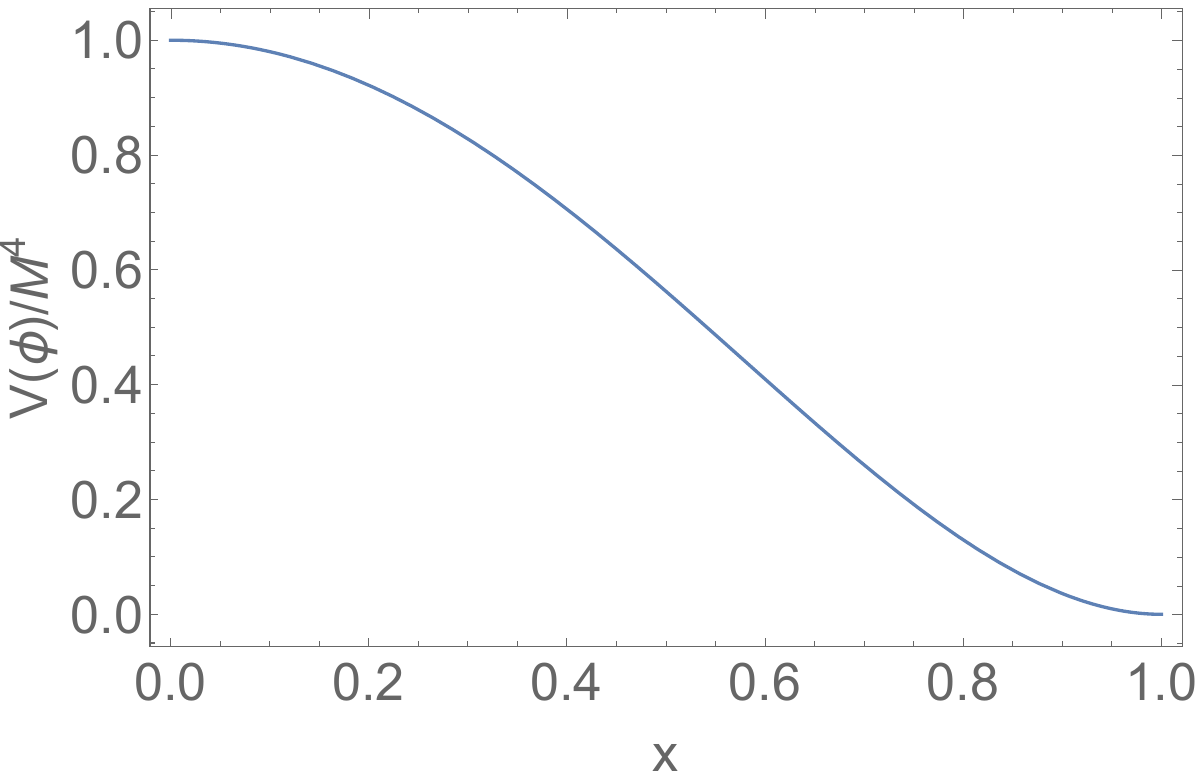}}
\caption{Double-Well potential as a function of $x$}
\end{figure}

With the potential in Eq. (\ref{12}) , the slow-roll parameters become,
\begin{align}
    \epsilon_V &= \frac{1}{1+ 2 \lambda} \frac{M_P^2}{\phi_0^2} \frac{8x^2}{(x^2 - 1)^2} \nonumber \\
    \eta_V &=  \frac{1}{1+ 2 \lambda} \frac{M_P^2}{\phi_0^2} \frac{12x^2 - 4}{(x^2 - 1)^2}
\end{align}

\begin{figure}[H]
\centerline{\includegraphics[width=0.5 \textwidth]{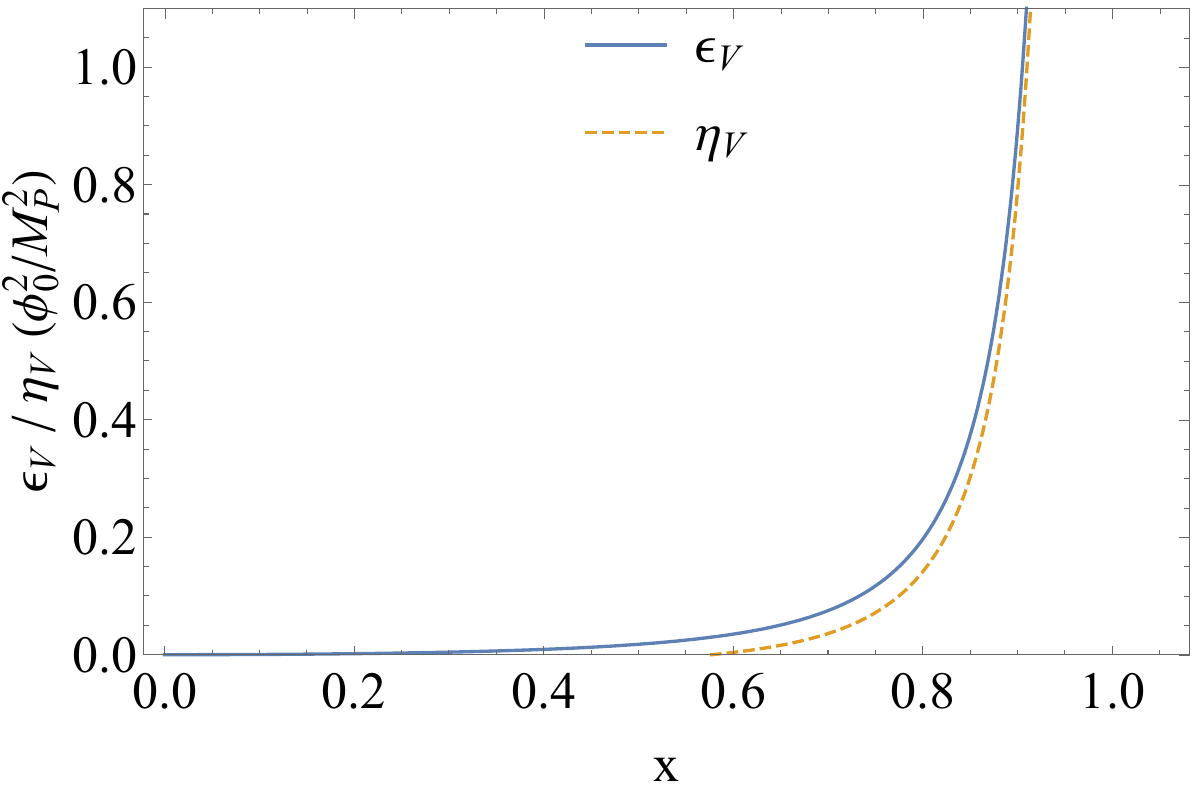}}
\caption{Plot for slow-roll parameters $\epsilon_V$ and $\eta_V$ as a function of $x$ for $\lambda=100$. The parameters are rescaled by the quantity $M_P^2/\phi_0^2$ , so that the corresponding expression becomes independent of $\phi_0$ }
\end{figure}

From the Fig.2, it is clear that $\epsilon_V$ is approaching the value 1 faster than $\eta_V$ and hence we will set $\epsilon_V$ equals to 1 to compute the value of the field at the end of inflation. From $\epsilon_V = 1$ we obtain the $x_{end}$ as,
\begin{equation}
    x_{end}= \frac{\frac{M_P}{\phi_0}}{\sqrt{1 + 2 \lambda}} \left[ - \sqrt{2} + \sqrt{2 + \left(\frac{\phi_0}{M_P}\right)^2 (1+2\lambda)} \right ]
    \label{14}
\end{equation}
Then from Eq. (\ref{13}) we obtain,
\begin{equation}
    N = (1+2\lambda)\left(\frac{\phi_0}{M_P}\right)^2 \left[ \frac{x^2}{8} - \frac{\ln x}{4} \right]^{x_i}_{x_{end}}
\end{equation}

Now setting $\phi_0=M_p$, on numerical calculation we found that for $\lambda=150$ and $x_i=0.28258$ we can get nearly $N=60$ no of e-folding. At this initial value of the inflaton, the corresponding scalar spectral index is $n_s= 0.9610$, tensor spectral index is $n_T= -0.00501$ and tensor-to-scalar ratio is $r=0.0401$ . This result is in good agreement with the Planck 2018 data \cite{r9} which sets the bound on scalar spectral index and tensor-to-scalar ratio as,
\begin{align}
     n_s = 0.9649 \pm 0.0042 \hspace{1cm} \text{(at 68 \% CL)} \\ \nonumber 
    r < 0.056 \hspace{2cm} \text{(at 95 \% CL)}
\end{align}
Now we will try to obtain the admissible range of the model parameter $\lambda$ for which the observable are in the range given in (35). For this, we vary the model parameter and obtain the below table.

\begin{table}[h!]
\centering
{\begin{tabular}{@{}ccccc@{}} \toprule
$\lambda$ & $x_i$  & $n_s$ & $r$ \\
 \midrule
145  &  0.2742   &  0.9606  &  0.0386 \\
146  &  0.27595   &  0.9607  &  0.0389 \\
221  &  0.3771   &  0.9649  &  0.0558 \\
222  &  0.3783   &  0.9650  &  0.0560 \\ \bottomrule
\end{tabular}
\caption{$n_S$ \& $r$ at different values of $\lambda$ for $N=60$ e-folds}
\label{ta1}}
\end{table}

Form the table we see that the admissible range of the model parameter is $145 < \lambda < 222$ for which $(n_s,r)$ stay within the Planck 2018 bounds. It is pertinent here to mention that for negative values of the model parameter $\lambda$, the solution of Eq. \ref{14} becomes imaginary for which physical interpretation isn't possible.


\section{Discussion and Conclusion}
Inflation has been the most accepted theory for describing the primordial universe. Recently, inflation has been studied in the frame work of ${f(R,T)}$ gravity with different potentials like quadratic potential, natural, hilltop potential and Starobinsky potential \cite{r13, r14, r15, r16}. Now, the functional form of ${f(R,T)}$ is a bit arbitrary and model parameters can be included for fine tuning to fit the observational data. In this work we have taken the simplest yet the most used form $f(R,T)=R+16\pi G \lambda T$ , where $\lambda$ is the model parameter. We reviewed the slow-roll inflation in the context of ${f(R,T)}$ gravity and it is found that correction due to the trace term has induced in all important equations. \\ \\
Then we introduced Double-Well potential and computed the cosmological observables for the model. We found that for $\lambda=150$ , the scalar spectral index is $n_s= 0.9610$ and tensor-to-scalar ratio is $r=0.0401$ which are in good agreement with Planck 2018 data. Further, it is seen that with the decrease in $\lambda$ , tensor-to-scalar ratio decreases. So, in future if further constraints will be imposed on r then the correction from ${f(R,T)}$ will help the model to fit the observational data. We then obtain the permissible range of the model parameter to be $145<\lambda<222$ in order to fit the Planck 2018 data. These results are valid for this model only. For other functional form of ${f(R,T)}$ result will vary.  \\ \\
Another important point to be discussed is that in case of Double-Well potential vacuum expectation value needs to be Super Planckian in order to match the results of the model with the observational data. But in our study we have set the vacuum expectation value at the Planck scale and obtain model results which are consistent with the observational CMB data. It implies that the inflaton vacuum expectation value in Double-Well inflation (DWI) can be reduced from Super Planckian to Planckian scale if we include a correction trace term in the theory. This ${f(R,T)}$ correction helps the DWI to be consistent with the Particle physics and effective field theory. \\ \\
Now as a possible extension of our work, one can study the same model for different other potentials. For a plethora of potential, one may check this reference \cite{r51}. Besides this, other forms of ${f(R,T)}$ as mentioned in the section 2 may be studied with different potentials. Further, Palatini and Metric-affine versions of ${f(R,T)}$ gravity may be checked to study inflation for different potentials which may generate interesting results.

\printbibliography

\end{document}